\documentclass[12pt]{article}
\usepackage{overpic}
\usepackage{pst-node}
\usepackage{epsfig,subfigure,wrapfig}
\begin{document}

\title{Senescence Can Explain Microbial Persistence}

\author{I. Klapper$^{a,b}$, P. Gilbert$^c$, B.P. Ayati$^d$, J. Dockery$^{a,b}$,
P.S. Stewart$^{b,e}$}

\date{}

\maketitle

\noindent
$^a$Department of Mathematical Sciences, Montana State University,
Bozeman, MT 59717\\
$^b$Center for Biofilm Engineering, Montana State University, Bozeman, MT
59717\\
$^c$School of Pharmacy and Pharmaceutical Sciences, University
of Manchester, Manchester, UK\\
$^d$Department of Mathematics, Southern Methodist University,
Dallas, TX 75205\\
$^e$Department of Chemical and Biological Engineering, Montana State University, Bozeman, MT 59717.
\vspace{.5in}

\noindent{Corresponding author: Isaac Klapper, Department of Mathematical
Sciences, Montana State University, Bozeman, MT 59717. Tel. (406)-994-5231.
Fax. (406)-994-1789. email: \verb+klapper@math.montana.edu+.}

\pagebreak

\begin{abstract}
It has been known for many years that small fractions of
persister cells resist killing in many bacterial colony-antimicrobial
confrontations. These persisters are not believed to be mutants. Rather it
has been hypothesized that they are phenotypic variants.
Current models allow cells to switch in and out of the
persister phenotype.
Here we suggest a different explanation, namely senescence,
for persister formation. Using a mathematical model
including age structure, we show that senescence provides
a natural explanation for persister-related phenomena
including the observations that persister fraction depends on
growth phase in batch culture and dilution rate in continuous
culture.
\end{abstract}
\vspace{.2in}

\noindent persisters $|$ senescence $|$ batch culture $|$ continuous culture
\vspace{.2in}

\section{Introduction}

It has been observed (Balaban et al., 2004; Bigger, 1944; 
Gilbert et al., 1990; Greenwood and O'Grady, 1970; Keren et al., 2004;
McDermott, 1958; Moyed and Bertrand, 1983; Sufya et al., 2003;
Haarison et al., 2005; Wiuff et al., 2005), dating to 
Bigger (1944), that many antimicrobials while effective in
reducing bacterial populations are unable to
eliminate them entirely, even with prolonged exposure. The surviving
cells, called persisters, may be small in number -- Bigger (1944)
reported less than 100 persisters out of $2.5\cdot10^7$ cells of 
{\it Staphylococcus pyogenes} after exposure to penicillin in some cases for
example -- but nevertheless are subsequently able upon
removal of the challenging agent to repopulate.
See Lewis (2001) for a general discussion. This phenomonon
has recently gained increased attention in the context of biofilms (Spoering
and Lewis, 2001)
where the persisting cells have the added protection of a polymeric matrix, 
making them particularly dangerous (see, e.g., models
of Roberts and Stewart, 2004, 2005).
The protection of microbial populations by persisters, whether formed by
senescence or some other mechanism, is expected to be enhanced in biofilms
because of the propensity of biofilms to harbor slow-growing or non-growing
cells.

Persisters have a number of interesting characteristics:

\begin{itemize}

\item
Upon reculturing, persister cells enable repopulation.
See, for example, Balaban et al. (2004).

\item
Persisters do not pass their tolerance to their progeny,
and progeny do not inherit any greater tendency to
be persisters. That is, persisters do not appear to
be genetic variants. See for example Keren et al. (2003)
and Balaban et al. (2004).

\item
It is observed that persister cells apparently grow slowly
or not at all in the presence of antimicrobial. See for example
Balaban et al. (2004).

\item Persister cells demonstrate resistance upon
exposure to multiple anti-microbial agents.
That is, persister cells with respect to one antimicrobial agent can also be
resistant upon exposure to different antimicrobial agents.
As an example see Sufya et al. (2003), where survivors (from an
{\it E. coli} batch culture) of a tetracycline challenge were
also tolerant to ciprofloxacin and quaternary ammonium compound.

\item
Bacterial cultures demonstrate biphasic killing patterns in
response to antimicrobial challenge. It has been suggested
that this plateauing is a consequence of the presence of persisters.
See for example Balaban et al. (2004).

\item 
In continuous culture experiments, persister fractions are
observed to increase with decreasing dilution rates, see
Sufya et al. (2003).

\item All other things being equal, a culture with an initially richer
growth medium will produce more persisters, see for example Bigger (1944).

\item
One of the more puzzling observations is that changes in the
population fraction of persister cells are growth-phase
dependent -- generally, though not always,
 population increases do not occur until the later
stages of the log phase or even until the stationary phase.
For example, Keren et al. (2004) observed an increasing
ratio of persister to total cell numbers over time
in {\it E. coli}, {\it P. aeruginosa}, and {\it S. aureus} batch cultures.
In Balaban et al. (2004) the authors went as far as to posit that differences in onset 
of persister cell increases between different strains of {\it E. coli}
imply the existence of more than one type of persister.

\end{itemize}
It been suggested that that persistence is a phenotypic phenomenon,
(e.g. Balaban et al. 2004; Cogan, 2006; Roberts and Stewart, 2005).
These references propose models in which cells are able to switch
in and out of a protected, slow- or non-growing persister state
with probablilities that are dependent possibly on environmental
conditions. In this paper, based on observations of microbial senescence
(Ackerman et al., 2003; Barker and Walmsley, 1999; Mortimer and Johnston, 1959;
Stewart et al., 2005), we instead propose
an alternative simple mechanism that can explain all of the above
mentioned properties.

We make only three essential assumptions:
\begin{enumerate}
\item bacterial cells age, \label{item1}
\item older cells are more resistant than younger cells
to antimicrobial challenge, and
\label{item2}
\item growth manifests as production of new cells. That is,
upon division one progenic cell inherits the effects of age while
the other does not. \label{item3}
\end{enumerate}
We thus regard older cells to be the persisters. In fact,
one might well label senescent cells to be a separate persister
phenotype (as opposed to a youthful phenotype) though we
do not stress this interpretation.

The second assumption could be a consequence for example of
decreased metabolic activity but the particular mechanism does not
really matter for our results. (The genetic bases for senescence
and persistence are just beginning to emerge; see for example
Nystrom, 2005, V\'{a}squez-Laslop et al., 2006, and Spoering et al., 2006.)
In reference to the third assumption,
we can interpret cell division as a new, youthful cell being
born from an old one.

We make the following additional non-essential specific assumptions
for definiteness:
\begin{enumerate}
\setcounter{enumi}{3}
\item production rate of new cells decreases with age but remains
greater than zero, \label{item4}
\item cell death occurs at a constant rate, and \label{item5}
\item a given concentration of applied antimicrobial will kill cells
of sufficiently young age but will not affect older cells. \label{item6}
\end{enumerate}
These extra assumptions matter in the details but do not affect the
qualitative results that we report with the exception that
a non-zero growth rate (in assumption~\ref{item4}) is necessary 
in order to enable persister cells to repopulate after antimicrobial
application. For consistency with the notion of decreasing activity,
we suppose that older cells grow more slowly than younger
cells although this assumption is again not really necessary here;
rather the focus is on senescence as a mechanism for resistance.

\section{Persister Model}\label{sec:modelling}

\subsection{Functional forms}

We define $b(t,a)\Delta a$ to be the size of the bacterial population 
between ages $a$ and $a+\Delta a$ at time $t$, and $c(t)$ to
be the growth media concentration at time $t$.
Following observations reported by Stewart et al. (2005), we
suppose that cells senesce at a linear rate in time; other
functional forms of age increasing  senescence could be
used instead.
In particular for a parameter $\lambda$ that we will 
call the {\it senescence time}, we define
$u(t,a)\Delta a$, the media usage
at time $t$ of cells between ages $a$ and $a+\Delta a$, by
\begin{displaymath}
u(c(t),a) = \left\{\begin{array}{ll}
 \alpha[(1-a/\lambda)+\xi]c(t), & a\leq\lambda,\\
 \alpha \xi c(t), & a>\lambda,
\end{array}\right.
\end{displaymath}
where $\alpha$ is a first-order rate constant and $\xi$
is a base usage factor.


Similarly,
we define $g(t,a)\Delta a$, the new cells born at time $t$
from cells between ages $a$ and $a+\Delta a$, by
\begin{displaymath}
g(c(t),a) = \beta u(c(t),a)
\end{displaymath}
where $\beta$ is a yield coefficient. 
Additionally, by assumption \ref{item5}, the death rate $\mu(a)$ takes the form
\begin{displaymath}
\mu(a) = \mu_0
\end{displaymath}
where $\mu_0$ is a death rate constant. Finally, functional
forms for antimicrobial application will be made below.

We stress that these choices are made for simplicity and
for consistency with available data from the literature.
The only essential condition we require is that the applied
antimicrobial agent exhibit less potency as cells age (see below).
For example, age-dependence in $u$ and $g$ is
unnecessary. Conversely, $\mu$ could be made functionally
dependent, for example, on $a$, $c(t)$ or $\lambda$ if so desired.

\subsection{Age Structure}

A mathematical description of age structure was first introduced
in Lotka (1907), McKendrick (1926), and many such representations
have been used since. 
Define $b(t,a)$ to be the bacterial population density of age $a$ at
time $t$, and $c(t)$ to be the amount of available media at time $t$. 
The equation governing $b$ is then
\begin{equation}
\frac{\partial b}{\partial t}(t,a) + \frac{\partial b}{\partial a}(t,a)
    =  -\mu_0 b(t,a)
    \label{b1eqn}
\end{equation}
where the term on the right-hand side reflects cell death. The
death coefficient $\mu_0$ can be expected, in general,
to depend on $c$ and $a$. We suppress such dependence here
for simplicity as it does not affect our results in a qualitative way.

Equation (\ref{b1eqn}) is valid for $a>0$. To obtain an equation
for $b(t,a=0)$, i.e., the new cells at time $t$, we observe that
such cells are ``born'' at time $t$ from the existing population
$b(t,a)$, $a>0$. For example, the subpopulation of cells between
ages $a$ and $a+\Delta a$ produces $g(c(t),a)b(t,a)\Delta a$
new cells. Summing then over the entire existing population
at time $t$, we obtain
\begin{equation}
b(t,0) = \int_{0}^{\infty}g(c(t),a)b(t,a)da.
    \label{b1init}
\end{equation}
A similar equation applies for the media concentration:
\begin{equation}
\frac{dc}{dt}(t) = -\int_{0}^{\infty}u(t,a)b(t,a)da.
    \label{c1eqn}
\end{equation}
Equations~(\ref{b1eqn})-(\ref{c1eqn}) are supplemented by initial conditions
$b(0,a)=b_0(a)$ for some supplied function $b_0(a)$,
and $c(0)=c_0$ for some supplied constant $c_0$. 
For discussion of mathematical issues involved
in age-differentiated systems such as the one
considered here, see e.g. Cushing (1998), Webb (1985).

We can write $u(t,a)=\alpha s(a)c(t)$ where
\begin{displaymath}
s(a) = \left\{\begin{array}{ll}
 1-a/\lambda + \xi, & a\leq\lambda,\\
 \xi, & a>\lambda,
\end{array}\right.
\end{displaymath}
($s(a)$ is the {\it senescence factor})
and so define a senescence-weighted total bacteria population
\begin{displaymath}
B(t)=\int_{0}^{\infty}s(a)b(t,a)da.
\end{displaymath}
Then (\ref{b1eqn})-(\ref{c1eqn}) become
\begin{eqnarray}
\frac{\partial b}{\partial t} + \frac{\partial b}{\partial a}
     & = & -\mu_0 b,\hspace{.2in}a>0 ,
    \label{b2eqn}\\
b(t,0) & = & \alpha\beta cB, 
    \label{b2init}\\
\frac{dc}{dt} & = & -\alpha cB,
    \label{c2eqn}
\end{eqnarray}
with initial conditions $b(0,a)=b_0(a)$, $c(0)=c_0$. In the
computations to follow we use $c_0=1$ kg/m$^3$ and
\begin{displaymath}
b_0(a) = \left\{\begin{array}{ll}
 (10^2{\rm\;cfu})\lambda^{-1}(1-a/2\lambda), & a\leq2\lambda,\\
 0{\rm\;cfu/hr}, & a>2\lambda.
\end{array}\right.
\end{displaymath}
With this choice, the total initial population is $10^2$ cfu distributed
linearly in age over the age interval $[0,2\lambda]$.

\subsection{Antimicrobial Application}

We include the effect of applied antimicrobial consistently with
assumption~\ref{item6}: a given
antimicrobial concentration $d$ applied at time $t$ to the bacteria
population results in killing at rate $\gamma$ of sufficiently
young (and hence susceptible) cells and does not effect older 
(and hence tolerant) cells. We assume for definiteness
and consistency with our senescence assumption that tolerance age
increases linearly with antimicrobial concentration. Other choices
of age dependence, as long as they are monotone in age, can
be made.

Then equation (\ref{b2eqn}) is replaced by
\begin{equation}
\frac{\partial b}{\partial t} + \frac{\partial b}{\partial a}
     = -(\mu_0+\mu_K(d,a))b,\hspace{.2in}a>0 ,
\label{b3eqn}
\end{equation}
where the rate $\mu_K$ is defined to be
\begin{equation}
\mu_K(d,a) = \left\{\begin{array}{cl}
 \gamma, & a\leq \delta d,\\
 0, & a>\delta d.
\end{array}\right.
\label{antimicrobial}
\end{equation}
Here $\delta$ is an adjustable tolerance coefficient.
For a given antimicrobial concentration $d$,
small $\delta$ means that cells become resistant at a 
relatively young age, and large $\delta$ means that cells
become resistant at a relatively old age. For numerical
reasons, we slightly smooth the discontinuity in $\mu_K$
in the results reported below.

Based on (\ref{antimicrobial}) then, for a given antimicrobial concentration $d$, cells of
age $\delta d$ or greater are designated as persisters. 

\subsection{Chemostat Model}

In addition to batch culturing, we consider a chemostat system
for which equations~(\ref{b2eqn})-(\ref{c2eqn}) become
\begin{eqnarray}
\frac{\partial b}{\partial t} + \frac{\partial b}{\partial a}
     & = & -(\mu_0+\tau)b,\hspace{.2in}a>0 ,
    \label{b2eqn_chemo}\\
b(t,0) & = & \alpha\beta cB, 
    \label{b2init_chemo}\\
\frac{dc}{dt} & = & -\alpha cB + \tau(C_0-c),
    \label{c2eqn_chemo}
\end{eqnarray}
where $\tau$ is the chemostat dilution rate and $C_0$ is the
reservoir substrate concentration.

\subsection{Parameters}

The model we describe contains seven parameters $\alpha$,
$\beta$, $\gamma$, $\delta$, $\lambda$, $\mu_0$, 
and $\xi$ (plus two, $\tau$ and $C_0$ for the chemostat), see Table~1.
\vspace{.4in}

\begin{tabular}{|l||l|l|}\hline
\multicolumn{3}{|c|}{Table 1: Variables and Parameters} \\ \hline
$B$ & weighted population at time $t$ & cfu \\ \hline
$b$ & population per age at age $a$ and time $t$ & cfu/hr \\ \hline
$C_0$ & reservoir substrate concentration & kg/m$^3$ \\ \hline
$c$ & substrate concentration at time $t$ & kg/m$^3$ \\ \hline
$d$ & antimicrobial dosage & kg/m$^3$ \\ \hline
$s$ & senescence factor & - \\ \hline
$\alpha$ & substrate usage rate & hr$^{-1}\cdot$cfu$^{-1}$ \\ \hline
$\beta$ & cell yield coefficient & cfu$\cdot$m$^3$/kg \\ \hline
$\gamma$ & cell killing rate & hr$^{-1}$ \\ \hline
$\delta$ & tolerance coefficient & hr$\cdot$m$^3$/kg \\ \hline
$\lambda$ & senescence time & hr \\ \hline
$\mu_0$ & nominal cell death rate & hr$^{-1}$ \\ \hline
$\tau$ & chemostat dilution rate & hr$^{-1}$ \\ \hline
$\xi$ & minimum substrate usage factor & - \\ \hline
\end{tabular}
\vspace{.4in}

In fact, the only essential constraints on these parameters 
necessary for the results we report are
that $\gamma>\mu_0$, i.e., the antimicrobial killing rate is greater
than the cell death rate, that $\delta>0$, i.e., cells become
more tolerant with age, and that $\xi>0$ so that persister cells
are capable of repopulation. 
We set the parameter values as follows:
\begin{itemize}
\item Growth and substrate usage parameters:
for definiteness, we set
the length of the log phase to be approximately 10 hr and the bacteria
doubling time to be approximately 0.75 hr, resulting in
approximately 13.3 doublings in the log phase. Given an initial
value of $10^2$ cfu, we thus obtain approximately $10^6$ cfu
at the end of the log phase. These constraints 
require $c_0\alpha\cong 2^{-13.3}\cong 10^{-6}$ and 
$c_0\alpha\beta\cong1.33=(0.75)^{-1}$
(in units as in Table 1). We set $c_0=1$ kg/m$^3$ as a reasonable
value for initial media concentration, and then use
$\alpha=10^{-6}$ hr$^{-1}$cfu$^{-1}$, $\beta=1.33\cdot10^6$ cfu$\cdot$m$^3$/kg.
(Roughly speaking, $\alpha^{-1}$ fixes the population at the end
of the log phase and $(\alpha\beta)^{-1}$ determines the length of the log
phase via the doubling time.)
The minimum substrate usage parameter $\xi$ is presumed to be small compared 
to 1, and
needs to be larger than zero in order for persister cells to repopulate.
Otherwise its value is unimportant. We set $\xi=10^{-3}$.
The value of the cell death rate is unimportant as well (and
could even be set to zero).
We use $\mu_0=0.05$ hr$^{-1}$.
\item Senescence parameter:
Supposing significant senescence
after about 16 generations (Stewart et al. (2005) reports
1-2\% per generation) with a cell division time of 0.75 hr,
we then obtain $\lambda=12$ hr. 
\item Antimicrobial parameters (except $\delta$):
As a typical antimicrobial dosage, we use $d=0.01$ kg/m$^3$.
As a typical antimicrobial killing rate, we use
$\gamma=10$~hr$^{-1}$. 
\item Chemostat parameters: we allow $\tau$, the dilution rate, to vary.
The other parameter, reservoir concentration $C_0$, does not
affect our results so is arbitrary.
\end{itemize}
The only important quantity without an estimate is the onset age
$\delta d$ at which bacteria become resistant to the
antimicrobial (see equation~(\ref{antimicrobial})). We assume that the 
onset age be equal to $\lambda$, the senescence time, and thus
$\delta=\lambda/d=1.5\cdot10^3$ hr$\cdot$m$^3$/kg. 
Increasing (decreasing) $\delta$
has the effect of increasing (decreasing) the onset age.

\subsection{Numerical Methods}

Equation (\ref{b3eqn}) along with (\ref{b2init}), (\ref{c2eqn})
are solved numerically.
We use a moving-grid Galerkin method in age with discontinuous
piecewise linear functions post-processed to cubic splines for the
approximation space in age (Ayati and Dupont, 2002). We use
a step-doubling method in time, Ayati and Dupont (2005).
This combination was illustrated in Section 5, Ayati and Dupont (2002),
using the same code as used for this paper.

\section{Results}

\begin{figure}
\includegraphics[height=75mm]{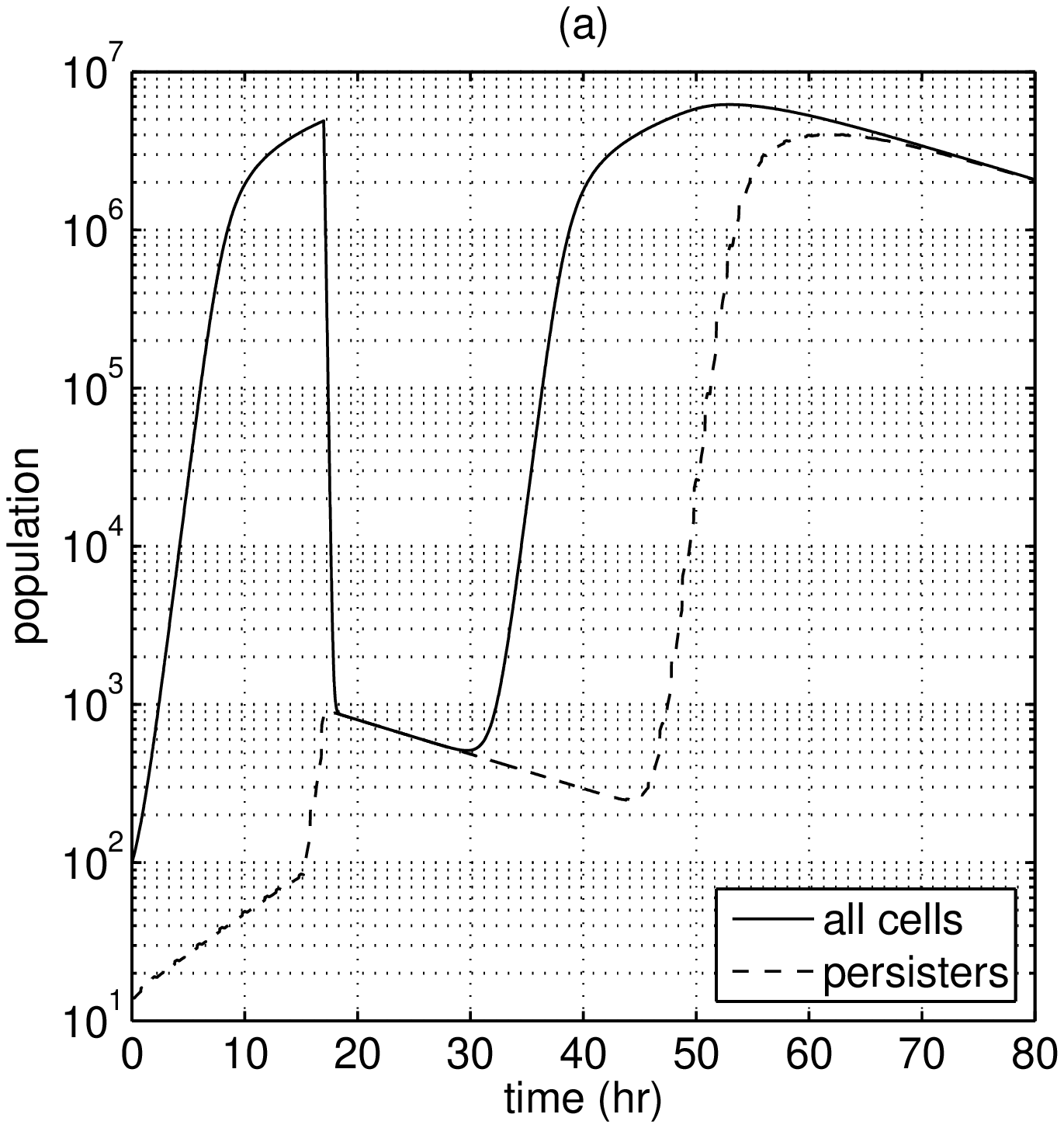}
\includegraphics[height=75mm]{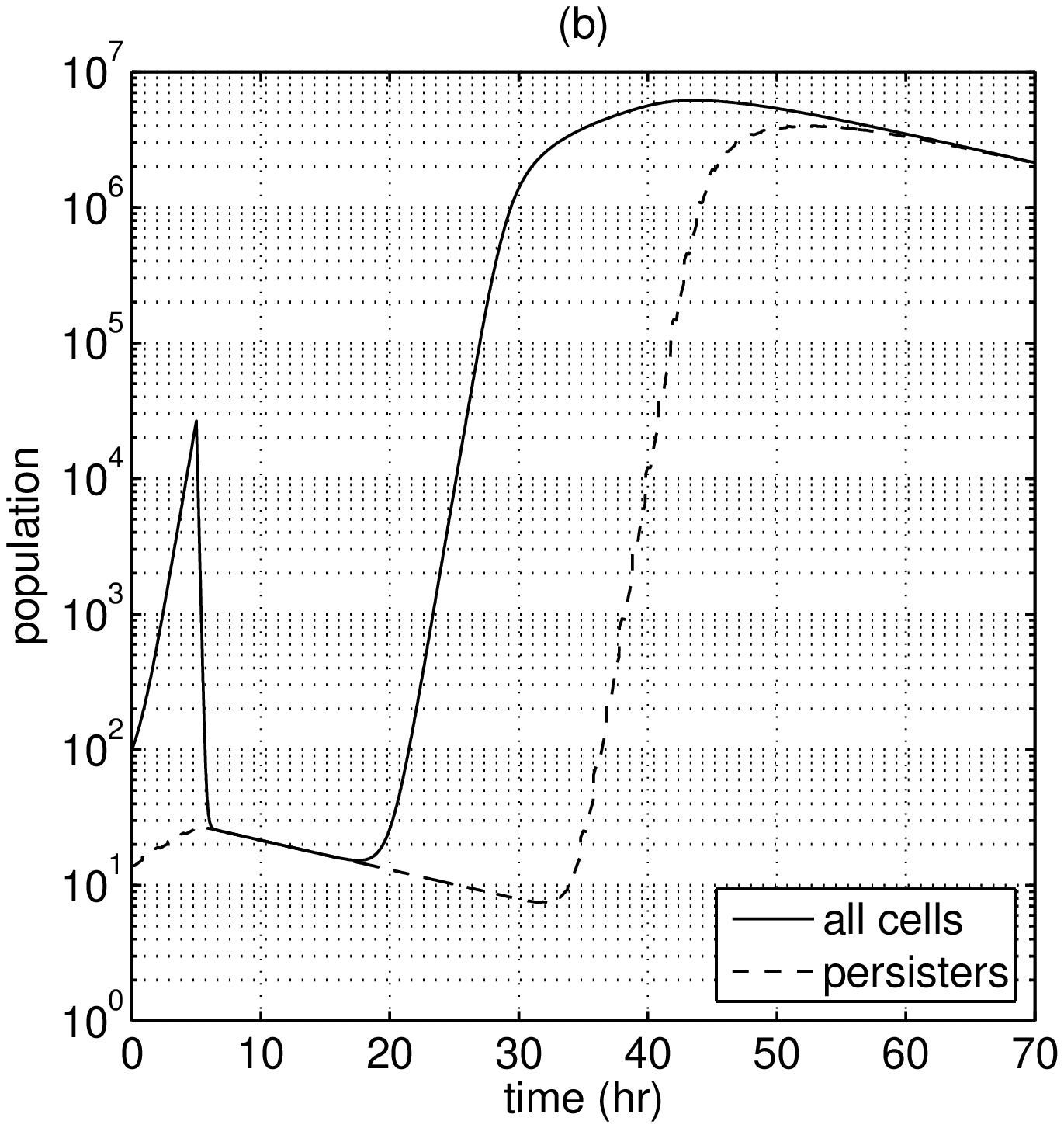}
\caption{Exposure of a batch culture to antimicrobial.
(a) antimicrobial is applied during stationary phase at $t=17$
and removed at $t=27$ at which time surviving cells are
recultured at the $t=0$ nutrient level.
(b) antimicrobial is applied during log phase at $t=5$ and removed
at $t=15$ at which time surviving cells are
recultured at the $t=0$ nutrient level.}
\label{persist_exper}
\end{figure}

In Figure \ref{persist_exper} we apply antimicrobial to a stationary
phase culture, Fig.~\ref{persist_exper}(a), and to a log phase
culture, Fig.~\ref{persist_exper}(b), and then ``reculture''
afterwards by removing the antimicrobial and adding fresh
medium. Note the biphasic survival
in each case: there is an initial sharp die-off immediately
after antimicrobial application followed by a second phase of
die-off of older (persister) cells due to natural causes.
We also note that the number of persisters 
in this example does not grow substantially
until the stationary phase -- this is a consequence of the
delay between birth of new cells in the log phase
and aging of those cells into persisters. That delay time
depends on $\delta$, our only free coefficient. Small $\delta$
results in short delay in the generation of persisters (i.e.,
growth in persister numbers even in the log phase) while
large $\delta$ results in long delay in the generation of
persisters. The growth at early times in persister numbers seen
in both Fig.~\ref{persist_exper}(a) and Fig.~\ref{persist_exper}(b) is
solely a consequence of our initial age distribution in
the innoculum. Note for example that persister numbers actually
decline during the second log phase in both 
Fig.~\ref{persist_exper}(a) and Fig.~\ref{persist_exper}(b).

\begin{figure}
\includegraphics[height=80mm]{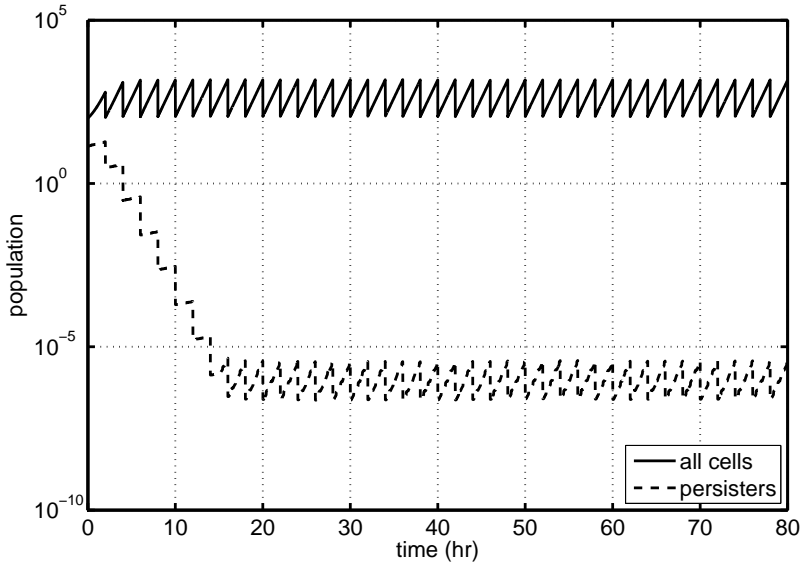}
\caption{Bacteria batch culture, diluted down to $10^2$
organisms every two hours. Solid curve is the total population
and dashed curve is the persister population.
See Keren et al. (2004), Fig. 4.}
\label{keren}
\end{figure}

In Figure \ref{keren} we present a computational version of the persister
elimination experiment conducted in Keren et al. (2004), see
Figure~4 of that reference, qualitatively matching the results
reported there. These authors observed that persister numbers
could be driven downwards by frequent reculturing. In addition
to serving as a comparison test of the persistence model to experiment,
our (computer) experiment serves to further emphasize
the point that the senescence mechanism presented here reproduces observational
evidence that persisters are not formed in early log phase, and
that by suppressing production of senescent cells, it is possible
to suppress persister frequency as in Kener et al. (2004).

\begin{figure}
\includegraphics[height=80mm]{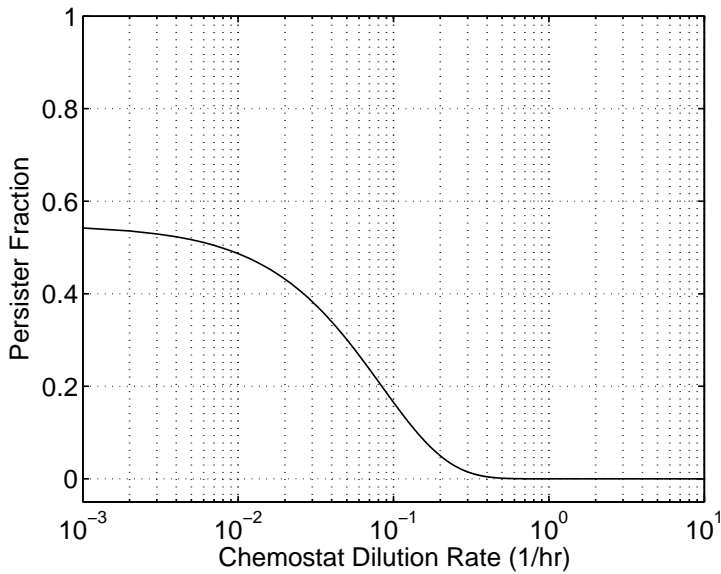}
\caption{Fraction of persisters in a steady state
chemostat as a function of chemostat dilution rate.}
\label{chemo}
\end{figure}

In fact, the persister elimination experiment illustrated
in Fig.~\ref{keren} might be considered as an approximation
to a chemostat with roughly 2 hour turnover. So
we also consider persister numbers in the chemostat 
system~(\ref{b2eqn_chemo})-(\ref{c2eqn_chemo}) at steady state
(see Sufya et al., 2003).
In particular, by setting $\partial b/\partial t=0$ in~(\ref{b2eqn_chemo})
and suppressing $t$ dependence, we obtain
\begin{displaymath}
b(a) = b(0)e^{-(\mu_0+\tau)a}.
\end{displaymath}
Thus the persister population fraction $P$ is given by
\begin{displaymath}
P = \frac{\int_{\delta d}^{\infty}b(a)\,da}{\int_{0}^{\infty}b(a)\,da}
  = e^{-(\mu_0+\tau)\delta d},
\end{displaymath}
see Figure \ref{chemo}. We again note the qualitative match
to the experimentally reported results in Sufya et al. (2003). 
In particular, note the rapid transition
from high to low persister fraction separating the two
regimes $\tau$ smaller than and larger than the persister
age $\delta d$. For small $\tau$, i.e., slow dilution, persister fraction tends to
a constant controlled by $\mu_0$, namely $e^{-\mu_0\delta d}\approx0.55$
for the parameter values used here. For large $\tau$, i.e., fast
dilution, persister fraction tends to zero. We remark that this
characterization of persister fraction with respect to dilution rate,
see Fig.~\ref{chemo},
is independent of the details of substrate usage. That is, it is
independent of equations (\ref{b2init_chemo}) and (\ref{c2eqn_chemo}).
It is also independent of the form of the senescence factor.
We note though that at sufficiently high dilution rate, washout of cells
exceeds the maximum specific growth rate and thus
the biomass concentration, including persisters, becomes zero
in steady state.

\section{Discussion}

The explanation of persistence as a symptom of senescence is
an attractive one. Tolerance (or resistance) due to age,
possibly because of reduced metabolic activity,
explains in a simple manner the persister characteristics listed in the
Introduction:
\begin{itemize}

\item Offspring of old cells are of course young, and able to
quickly repopulate.

\item Offspring of old cells are young and metabolically
active cells, and hence do not inherit low activity tolerance,
i.e., are not themselves persisters.

\item Old cells grow slowly.

\item Metabolic activity is independent of choice of antimicrobial agent
so that metabolically inactive cells can demonstrate resistance
to multiple agents.

\item In the presence of an antimicrobial agent, non-persisters are killed
quickly followed by slow die-off of persisters, i.e.,
biphasic behavior occurs.

\item 
Decreasing the dilution rate in a continuous culture allows
more cells to age longer before washout, hence increasing
the persister fraction.

\item Increasing the amount of initial growth medium allows
more new cells to be generated and hence, once these new
cells have aged, more persisters.

\item Increase in persister numbers can appear to depend on
growth phase. In particular new cells are, obviously,
young cells and so log phase growth does not affect
persister numbers until enough time has passed for those
new cells to age sufficiently. Hence persister numbers
do not significantly increase until later in the log phase or even
the stationary phase (and may actually decrease before).

\end{itemize}

All of these observations are qualitative properties of
our model depending on assumptions~(\ref{item1})-(\ref{item3})
and, we believe, essentially independent of the particular choices made
in the other assumptions.


A number of authors have previously suggested persisters
to be switching phenotypic variants  (e.g. Balaban et al., 2004, Cogan, 2006, 
Kussell et al., 2005, Roberts and Stewart, 2005, Sufya et al., 2005,
Wiuff et al., 2005), that is, that persisters are cells with the same
genome but with different sets of genetic expression as ``normal''
cells, and that a given cell can switch back and forth
between the two states. The resulting phenotype switching
model consists then of cells transiting between persister
and non-persister phenotypes. Available data does not distinguish between
the alternative persister explanations of senescence and
phenotypic switching. However, phenotype switching models
require transitions between phenotypes to be growth phase
dependent in order to explain phase dependence of persister
numbers -- persisters are more abundant in late stationary
phase than in earlier stages. Hence, it becomes
necessary to introduce stage-dependent transition rates,
perhaps by making those rates nutrient dependent. 
Further, observations indicate that
persistence in a given colony is antimicrobial dependent, that is,
the number of surviving persister cells can depend on the choice
of antimicrobial. Thus a phenotype switching model would
seem to require a special parameter set (indicating, possibly,
a special persister) for each
distinguished antimicrobial. On the other
hand, the senescence model can handle this issue simply by 
allowing for different antimicrobials
to be effective up until different levels of senescence, amounting
in our presentation to making the parameter $\delta$ 
be antimicrobial dependent. So, while we are unable yet to
distinguish between phenotypic and senescent explanations
of persisters based on available data, we can argue that
the mechanism of senescence provides a simpler explanation
at least with regards to the phenomenon discussed in
this paragraph.

We remark in conclusion that since 
(i) the segregation of chromosomes between mother and daughter 
cell is non-random and favors the parent with respect to the original rather 
than the copy strand, and (ii) mutations accumulate with successive 
replications of the chromosome then senescence/persistence/conservation 
of the older cells, particularly at times of antimicrobial stress, 
will ensure that following catastrophic stress of a population, 
repopulation will be from older, demonstrably successful ``archive'' cells. 
This would avoid evolutionary dog-legs and possible extinction.
Furthermore, consequent reliance upon the younger,
mutation-prone cells to populate the community would increase
the chance of adventitous genotypes being available in times
of stress, whilst maintaining the ``fall-back'' position
of reincarnation for the archived persisters.

\section{Acknowledgements}
I.K., J.D., and P.S. would like to acknowledge support from 
NIH award 5R01GM67245. I.K. and B.A. would like to thank IPAM,
where much of this work was conducted, for its hospitality.

\pagebreak


\section{Bibliography}


\noindent
Ackerman, M., Stearns, S.C., Jenal, U., 2003.
Senescence in a bacterium with asymmetric division,
{\it Science}, {\bf 300} 1920.\\
\vspace{.2in}

\noindent
Ayati, B.P., Dupont, T.F., 2002.
\uppercase{G}alerkin methods in age and space for a 
population model with nonlinear diffusion,
{\it SIAM J. Numer. Anal.}, {\bf 40} 1064-1076.\\
\vspace{.2in}

\noindent
Ayati, B.P., Dupont, T.F., 2005.
Convergence of a step-doubling \uppercase{G}alerkin 
method for parabolic problems,
{\it Math. Comp.}, {\bf 74} 1053-1065.\\
\vspace{.2in}

\noindent
Balaban, N.Q., Merrin, J., Chait, R., Kowalik, L., Leibler, S., 2004.
Bacterial persistence as a phenotypic switch, 
{\it Science}, {\bf 305} 1622-1625.\\
\vspace{.2in}

\noindent
Barker, M.G., Walmsley, R.M., 1999.
Replicative ageing in the fission yeast {\it Schizosaccharomyces pombe},
{\it Yeast}, {\bf 15} 1511-1518.\\
\vspace{.2in}

\noindent
Bigger, J.W., 1944.
{\it Lancet}, {\bf ii}:497-500.\\
\vspace{.2in}

\noindent
Cogan, N.G., 2006.
Effects of persister formation on bacterial response to dosing,
{\it J. Theor. Biol.}, {\bf 3} 694-703.\\
\vspace{.2in}

\noindent
Cushing, J.M., 1998. {\it An Introduction to Structured Population Dynamics},
SIAM, Philadelphia.\\
\vspace{.2in}

\noindent
Gilbert, P., Collier, P.J., Brown, M.R.W., 1990.
Influence of growth rate on susceptibility to antimicrobial agents:
biofilms, cell cycle, dormancy, and stringent response,
{\it Antimicrob. Agents Chemother.}, {\bf 34} 1865-1868.\\
\vspace{.2in}

\noindent
Greenwood, D., O'Grady, F., 1970.
Trimodal response of {\it Escherichia coli} and {\it Proteus Mirabilis}
to penicillins,
{\it Nature}, {\bf 228} 457-458.\\
\vspace{.2in}

\noindent
Harrison, J.J., Ceri, H., Roper, N.J., Badry, E.A., Sproule, K.M., 
Turner, R.J., 2005,
Persister cells mediate tolerance to metal oxyanions in
{\it Escherichia coli}, {\it Microbiology}, {\bf 151} 3181-3195.
\vspace{.2in}

\noindent
Keren, I., Shah, D., Spoering, A., Kaldalu, N., Lewis, K. (2004)
Specialized persister cells and the mechanism of multidrug
tolerance in {\it Escherichia coli},
{\it J. Bacteriol.}, {\bf 186} 8172-8180.\\
\vspace{.2in}


\noindent
Kussell, E., Kishony, R., Balaban, N.Q., Leibler, S. (2005)
Bacterial persistence: a model of survival in changing environments, 
{\it Genetics}, {\bf 169} 1807-1814.\\
\vspace{.2in}

\noindent
Lewis, K., 2001.
Riddle of biofilm resistance,
{\it Antimicrob. Agents Chemother.}, {\bf 45} 999-1007.\\
\vspace{.2in}

\noindent
Lotka, A.J., 1907.
Studies on the mode of growth of material aggregates,
{\it American J. Science}, {\bf 24} 141-158.\\
\vspace{.2in}

\noindent
McDermott, W., 1958.
 Microbial Persistence,
{\it Yale J. Biol. Med.}, {\bf 30} 257-291.\\
\vspace{.2in}

\noindent
McKendrick, A.G., 1926.
Applications of mathematics to medical problems,
{\it Proc. Edin. Math. Soc.}, {\bf 44} 98-130.\\
\vspace{.2in}

\noindent
Mortimer, R.K., Johnston, J.R., 1959.
Life span of individual yeast cells, 
{\it Nature}, {\bf 183} 1751-1752.\\
\vspace{.2in}

\noindent
Moyed, H.S., Bertrand, K.P., 1983.
hipA, a newly recognized gene of {\it Escherichia coli} K-12 
that affects frequency of persistence after inhibition of murein synthesis,
{\it J. Bacteriol.}, {\bf 155} 768-775.\\
\vspace{.2in}

\noindent
Nystrom, T. 2005. 
Bacterial senescence, programmed death, and premeditated
sterility. {\it ASM News}, {\bf 71} 363-369.\\
\vspace{.2in}

\noindent
Roberts, M.E., Stewart, P.S., 2004.
Modeling antibiotic tolerance in biofilms by accounting for nutrient limitation,
{\it Antimicrob. Agents Chemother.}, {\bf 48} 48-52.\\
\vspace{.2in}

\noindent
Roberts, M.E., Stewart, P.S., 2005.
Modelling protection from antimicrobial agents in biofilms
through the formation of persister cells,
{\it Microbiology}, {\bf 151} 75-80.\\
\vspace{.2in}

\noindent
Spoering, A.L., Lewis, K., 2001.
Biofilms and planktonic cells of {\it Pseudomonas aeruginosa} 
have similar resistance to killing by antimicrobials, 
{\it J. Bacteriol.},  {\bf 183} 6746-6751.\\
\vspace{.2in}

\noindent
Spoering, A.L., Vulic, M., Lewis, K., 2006.
{\it GlpD} and {\it PlsB} participate in
persister cell formation in {\it Escherichia coli},
{\it J. Bacteriol.},  {\bf 188} 5136-5144.\\
\vspace{.2in}

\noindent
Stewart, E.J., Madden, R., Paul, G., Taddei, F., 2005.
Aging and death in an organism that reproduces by morphologically
symmetric division, 
{\it PLoS Biology}, {\bf 3} 295-300 (2005).\\
\vspace{.2in}

\noindent
Sufya, N., Allison, D.G., Gilbert, P., 2003.
Clonal variation in maximum specific growth rate and
susceptibility towards antimicrobials, 
{\it J. Appl. Microbiol.} {\bf 95}, 1261-1267.\\
\vspace{.2in}

\noindent
V\'{a}zquez-Laslop, N., Lee, H., Neyfakh, A., 2006.
Increased persistence in {\it Escherichia coli} caused by controlled
expression of toxins and other unrelated proteins.
{\it J. Bacteriol.}, {\bf 188} 3493-3497.
\vspace{.2in}

\noindent
Wiuff, C., Zappala, R.M., Regoes, R.R., Garner, K.N., Baquero, F., 
Levin, B.R., 2005.
Phenotypic tolerance: antibiotic enrichment of noninherited 
resistance in bacterial populations,
{\it Antimicrob. Agents Chemother.}, {\bf 49} 1483-1494.\\
\vspace{.2in}

\noindent
Webb, G.F., 1985. {\it Theory of Nonlinear Age-Dependent Population Dynamics},
Marcel Dekker, Inc., New York.\\
\vspace{.2in}


\end{document}